\documentclass[aps,twocolumn]{revtex4}
\usepackage{epsfig}
\begin{document}
\renewcommand{\thefigure}{\arabic{figure}}
\setcounter{figure}{0}

\bibliographystyle{apsrev}

\title{Tests of Bayesian Model Selection Techniques for Gravitational Wave Astronomy}

\author{ \surname{Neil} J. Cornish and \surname{Tyson} B. Littenberg}
\affiliation{Department of Physics, Montana State University, Bozeman,
MT 59717}

\begin{abstract}
The analysis of gravitational wave data involves many model selection
problems. The most important example is the detection problem of selecting
between the data being consistent with instrument noise alone, or instrument
noise and a gravitational wave signal. The analysis of data from
ground based gravitational wave detectors is mostly conducted
using classical statistics, and methods such as the Neyman-Pearson
criteria are used for model selection. Future space based detectors, such
as the \emph{Laser Interferometer Space Antenna} (LISA), are expected
to produced rich data streams containing the signals from many millions
of sources. Determining the number of sources that are resolvable, and
the most appropriate description of each source poses a challenging model
selection problem that may best be addressed in a Bayesian framework.
An important class of LISA sources are the millions of
low-mass binary systems within our own galaxy, tens of thousands of
which will be detectable. Not only are the number of sources unknown, but so are the
number of parameters required to model the waveforms. For example, a significant
subset of the resolvable galactic binaries will exhibit orbital
frequency evolution, while a smaller number will have measurable
eccentricity. In the Bayesian approach to model selection one needs
to compute the Bayes factor between competing models. Here we explore
various methods for computing Bayes factors in the context of
determining which galactic binaries have measurable frequency evolution.
The methods explored include a Reverse Jump Markov Chain Monte Carlo
(RJMCMC) algorithm, Savage-Dickie density ratios, the Schwarz-Bayes
Information Criterion (BIC), and the Laplace approximation to the model
evidence. We find good agreement between all of the approaches.
\end{abstract}

\maketitle

\section{Background} 
Bayesian statistical techniques are becoming increasingly popular in gravitational wave
data analysis, and have shown great promise in tackling the
various difficulties of gravitational wave (GW) source extraction from
modeled data for the Laser Interferometer Space Antenna (LISA).
A powerful tool in the suite of Bayesian methods
is that of quantitative model selection~\cite{bayes2,bayes}.
To understand why this is a valuable feature consider a scenario where
one is attempting to fit data with two competing models of differing
dimension.  In general, a higher dimensional model will produce
a better fit to a given set of data.  This can be taken to the limit
where there are as many model parameters as there are data points allowing
one to perfectly match the data.  The problem then is to
decide how many parameters are physically meaningful and to select the
model containing only those parameters.  In the
context of GW detection these
extra parameters could be additional physical parameters used to
model the source or additional sources in the data.  If a model is
over-parameterized it will over-fit the data and produce spurious results.

Many of the model selection problems associated with LISA astronomy involve
{\em nested} models, where the simpler model forms a subset of the
more complicated model. The problem of determining the number of resolvable
galactic binaries, and the problem of determining the number of measurable
source parameters, are both examples of nested model selection. One could
argue that the later is better described as ``approximation selection'' since
we are selecting between different parameterizations of the full 17 dimensional
physical model that describes the signals from binary systems of
point masses in general relativity. However, many similar modeling problems in astrophysics
and cosmology~\cite{bayes}, as well as in other fields such
as geophysics~\cite{RJ}, are considered to be examples of model selection, and
we will adopt that viewpoint here.

The LISA observatory~\cite{lppa} is designed to explore the low frequency portion
of the gravitational wave spectrum between $\sim 0.1 \rightarrow 100$ mHz. This
frequency region will be heavily populated by signals from galactic binary systems composed
of stellar mass compact objects (e.g. white dwarfs and
neutron stars), of which millions are theorized to exist.  Tens of
thousands of these GW sources will be resolvable by LISA and the
remaining sources will contribute to a confusion-limited background~\cite{seth}.
This is expected to be the dominant source of low frequency noise for
LISA. 

Detection and subsequent regression of the galactic foreground is of vital
importance in order to then pursue dimmer
sources that would otherwise be buried by the foreground.
Because of the great number of galactic sources, and the ensuing
overlap between individual sources, a one-by-one detection/regression
is inaccurate~\cite{MCMC2}.  Therefore a global
fit to all of the galactic sources is required. Because of the
uncertainty in the number of resolvable sources one can
not fix the model dimension \emph{a priori} which presents a
crucial model selection problem.  Over-fitting the data will result in
an inaccurate regression which would then remove power from other
sources in the data-stream, negatively impacting their detection
and characterization. The Reverse Jump Markov Chain Monte Carlo approach to
Bayesian model selection has been used to determine the number of resolvable
sources in the context of a toy problem~\cite{andrieu, woan2} which shares
some of the features of the LISA foreground removal problem.  Meanwhile the Laplace
approximation to Bayesian model selection has been employed to estimate
the number of resolvable sources as part of a MCMC based algorithm to
extracting signals from simulated LISA data streams~\cite{MCMC2,BAM}.

Another important problem for GW astronomy is the determination
of which parameters need to be included in the description of the waveforms.
For example, the GW signal from a binary inspiral, as detected by LISA,
may involve as many as 17 parameters. However, for massive black hole binaries
of comparable mass we expect the eccentricity to be negligible, reducing the model dimension to
15, while for extreme mass ratio systems we expect the spin of the smaller body
to have little impact on the waveforms, reducing the model dimension to
14. In many cases the inspiral signals may be described by even fewer parameters.
For low mass galactic binaries spin effects will be negligible (removing six parameters),
and various astrophysical processes will have circularized the orbits of the majority of
systems (removing two parameters). Of the remaining nine parameters, two describe
the frequency evolution - {\it e.g.} the first and second time derivatives of the
GW frequency, which may or may not be detectable~\footnote{While this count is
only strictly correct for point-like masses, frequency evolution due to tides and
mass transfer can also be described by the same two parameters for the majority of
sources in the LISA band.}. 

Here we investigate the application of Bayesian model selection to LISA data analysis
in the context of determining the conditions under which the first time derivative
of the GW frequency, $\dot{f}$, can be inferred from the data. We parameterize the
signals using the eight parameters
\begin{equation}
\vec{\lambda}\rightarrow(A, f, \theta, \phi, \psi, \iota, \varphi_0, \dot{f})
\end{equation}
where $A$ is the amplitude, $f$ is the initial gravitational wave
frequency, $\theta$ and $\phi$ are the ecliptic co-latitude and
longitude, $\psi$ is the polarization angle, $\iota$ is the orbital
inclination of the binary and $\varphi_0$ is the GW phase.  The parameters
$f$, $\dot{f}$ and $\varphi_0$ are evaluated at some fiducial time ({\it e.g.}
at the time the first data is taken). For our analysis only a single source is
injected into the simulated data streams. In the frequency domain the output $s(f)$
in channel $\alpha$ can be written as    
\begin{equation} 
\tilde{s}_{\alpha}(f) = \tilde{h}_{\alpha}(\vec{\lambda}) + \tilde{n}_{\alpha}(f)
\end{equation}
where $\tilde{h}_\alpha(\vec{\lambda})$ is the response of the detector to
the incident GW and $\tilde{n}_{\alpha}(f)$ is the instrument noise.
For our work we will assume stationary Gaussian instrument noise with no
contribution from a confusion background. In our analysis we use the
noise orthogonal $A,E,T$ data streams, which can be constructed from linear
combinations of the Michelson type $X,Y,Z$ signals:
\begin{eqnarray}
A = \frac{1}{3}\left(2 X - Y -Z\right), \nonumber \\
E = \frac{1}{\sqrt{3}}\left(Z - Y\right), \nonumber \\
T = \frac{1}{3}\left(X + Y + Z\right).
\end{eqnarray}
This set of $A,E,T$ variables differ slightly from those
constructed from the Sagnac signals~\cite{aet}. We do not use
the $T$ channel in our analysis as it is insensitive to GWs at the frequencies
we are considering. The noise spectral density in the $A,E$ channels has
the form
\begin{eqnarray}
&& S_n(f) = \frac{4}{3}\left(1-\cos(2f/f_*)\right)
\Big((2+\cos(f/f_*))  S_s  \nonumber \\
&&  \quad + 2(3+2\cos(f/f_*)+\cos(2f/f_*))S_a\Big)\; {\rm Hz}^{-1}
\end{eqnarray}
where $f_*= 1/2\pi L$, and the acceleration noise $S_a$ and shot noise $S_s$ are simulated
at the levels
\begin{eqnarray}
S_a &=& \frac{10^{-22}}{L^2} \; {\rm Hz}^{-1}\nonumber\\
S_s &=& \frac{9\times10^{-30}}{(2\pi f)^4 L^2} \; {\rm Hz}^{-1} \, .\nonumber\\
\end{eqnarray}
Here $L$ is the LISA arm length ($\approx 5\times10^9$ m).

Of central importance to Bayesian analysis is the posterior distribution function (PDF) of the
model parameters.  The PDF $p(\vec{\lambda}|s)$ describes the probability that the source is
described by parameters $\vec{\lambda}$ given the signal $s$. According to
Bayes' Theorem,
\begin{equation}\label{posterior}
p(\vec{\lambda}|s)=\frac{p(\vec{\lambda}) p(s|\vec{\lambda})}
{\int d\vec{\lambda}\  p(\vec{\lambda}) p(s|\vec{\lambda})}
\end{equation}
where $p(\vec{\lambda})$ is the \emph{a priori}, or prior,
distribution of the parameters $\vec{\lambda}$ and $p(s|\vec{\lambda})$
is the likelihood that we measure the signal $s$ if the source is described
by the parameters $\vec{\lambda}$. We define the likelihood using the noise
weighted inner product 
\begin{equation}\label{nwip}
(A|B)\equiv \frac{2}{T} \sum_{\alpha} \sum_f
  \frac{\tilde{A}_{\alpha}^{\ast}(f)\tilde{B}_{\alpha}(f)+\tilde{A}_{\alpha}(f)\tilde{B}_{\alpha}^{\ast}(f)}{S_n^{\alpha}(f)} 
\end{equation}
 as 
\begin{equation}\label{like}
p(s|\vec{\lambda})=C\exp\Big[-\frac{1}{2}\Big(s-h(\vec{\lambda})\Big | s-h(\vec{\lambda})\Big)\Big]
\end{equation}
where the normalization constant $C$ depends on the noise, but not the GW signal.
One goal of the data analysis method is to find the parameters $\vec{\lambda}$
which maximizes the posterior.  Markov Chain Monte Carlo (MCMC)
methods are ideal for this type of application~\cite{MCMC1}.  The MCMC algorithm will simultaneously
find the parameters which maximize the posterior and accurately
map out the PDF of the parameters.  This is achieved through the use
of a Metropolis-Hastings~\cite{metro, haste} exploration of the parameter
space.  A brief description of this process is as follows:  The chain begins at some random
position $\vec{x}$ in the parameter space and subsequent steps are made by
randomly proposing a new position in the parameter space $\vec{y}$.
This new position is determined by
drawing from some proposal distribution $q(\vec{x}|\vec{y})$.
The choice of whether or not adopt the new position $\vec{y}$ is made by calculating
the Hastings ratio (transition probability)
\begin{equation}
\alpha = \min
\Big{\{}1,\frac{p(\vec{y}) p(s|\vec{y}) q(\vec{y}|\vec{x})}
{p(\vec{x}) p(s|\vec{x}) q(\vec{x}|\vec{y})} \Big{\}}
\end{equation}
and comparing $\alpha$ to a random number $\beta$ taken from a uniform draw in the interval [0:1].
If $\alpha$ exceeds $\beta$ then the chain adopts $\vec{y}$ as the
new position.  This process is repeated until some convergence
criterion is satisfied.  The MCMC differs from a Metropolis extremization by forbidding proposal
distributions that depend on the past history of the chain. This ensures that the progress of the
chain is Markovian and therefore statistically unbiased.  Once the chain has stabilized in the
neighborhood of the best fit parameters all previous steps of the chain
are excluded from the statistical analysis (these early
steps are referred to as the ``burn in'' phase of the chain) and
henceforth the number of iterations the chain spends
at different parameter values can be used to infer the PDF.  

The power of the MCMC is two-fold:  Because the algorithm has a
finite probability of moving away from a favorable location in the
parameter space it can avoid getting trapped by local features of the
likelihood surface.  Meanwhile, the absence of any ``memory'' within the
chain of past parameter values allows the algorithm to blindly,
statistically, explore the region in the neighborhood of the global
maximum.  It is then rigorously proven that an MCMC will (eventually)
perfectly map out the PDF, completely removing the need for user input
to determine parameter uncertainties or thresholds.

The parameter vector that maximizes the posterior
distribution is stored as the maximum \emph{a posteriori} (MAP) value
and is considered to be the best estimate of the source parameters.
Note that because of the prior weighting in the definition of the PDF
this is not necessarily the $\vec{\lambda}$ that yields the greatest
likelihood.  Upon obtaining the MAP value for a particular model $X$
the PDF, now written as $p(\vec{\lambda,X}|s)$, can be employed to solve the model selection problem.

\section{Bayes Factor Estimates}
The Bayes Factor $B_{XY}$ \cite{Jeffreys} is a comparison of the \emph{evidence} for two competing models, $X$ and $Y$, where
\begin{equation}\label{evidence}
p_{X}(s)=\int d\vec{\lambda}\ p(\vec{\lambda},X|s)
\end{equation}
is the marginal likelihood, or evidence, for model X.  The Bayes Factor can then be calculated by
\begin{equation}
B_{XY}(s)=\frac{p_X(s)}{p_Y(s)}.
\end{equation}
The Bayes Factor has been described as the Holy
Grail of model selection:  It is a powerful entity that is very difficult to find.
The quantity $B_{XY}$ can be thought of as the odds ratio for a preference of
model $X$ over model $Y$.
\begin{table}[h]
\begin{tabular}{|c|c|c|}
\hline
$B_{XY}$ & $2\log{B_{XY}}$ & Evidence for model $X$\\
\hline 
$<$ 1 & $<$ 0 & Negative (supports model $Y$)\\
1 to 3 & 0 to 2 & Not worth more than a bare mention\\
3 to 12 & 2 to 5 & Positive\\
12 to 150 & 5 to 10 & Strong\\
$>$ 150 & $>$ 10 &  Very Strong\\
\hline  
\end{tabular}
\label{tab1}
\caption{$B_{XY}$ `confidence' levels taken from~\cite{bayes2}}
\end{table}
Apart from carefully concocted toy problems, direct calculation of the
evidence, and therefore $B_{XY}$, is impractical.  To determine $B_{XY}$ the integral
required to compute $p_X$ can not generally be solved analytically and
for high dimension models Monte-Carlo integration proves to be
inefficient.  To employ this powerful statistical tool
various estimates for the Bayes Factor have been developed that
have different levels of accuracy and computational cost~\cite{bayes2, bayes}.
We have chosen to focus on four such methods: Reverse Jump Markov
Chain Monte Carlo and
Savage-Dickie density ratios, which directly estimate the Bayes
factor, and the Schwarz-Bayes Information Criterion (BIC) and
Laplace approximations of the model evidence.

\subsection{RJMCMC}
Reverse Jump Markov Chain Monte Carlo (RJMCMC) algorithms are a class of MCMC
algorithms which admit ``trans-dimensional'' moves between models of different
dimension~\cite{Green, RJLMBIC, RJ}.  For the trans-dimensional implementation applicable to the
LISA data analysis problem the choice of model parameters becomes
one of the search parameters.  The algorithm proposes parameter `birth'
or `death' moves (proposing to include or discard the
`extra' parameter(s)) while holding all other parameters fixed.  The
priors in the RJMCMC Hastings ratio
\begin{equation}\label{modhastings}
\alpha = \min
\Big{\{}1,\frac{p(\vec{\lambda}_Y) p(s|\vec{\lambda}_Y) g(\vec{u}_Y)}
{p(\vec{\lambda}_X) p(s|\vec{\lambda}_X) g(\vec{u}_X)}|\bf{J}| \Big{\}}
\end{equation}
automatically penalizes the posterior density of the higher
dimensional model, which compensate for its generically higher
likelihood, serving as a built in `Occam Factor.'  The $g(\vec{u})$
which appears in (\ref{modhastings}) is the distribution from which the
random numbers $\vec{u}$ are drawn and $|\bf{J}|$ is the Jacobian
\begin{equation}
|{\bf J}| \equiv \Big |
 \frac{\partial(\vec{\lambda}_Y,\vec{u}_Y)}{\partial(\vec{\lambda}_X,\vec{u}_X)}
 \Big |
\end{equation}.

The chain will tend to spend more iterations using
the model most appropriately describing the data, making the
decision of which model to favor a trivial one.  To quantitatively determine the
Bayes Factor one simply computes the ratio of the iterations
spent within each model.
\begin{equation}
B_{XY}\simeq \frac{\text{\# of iterations in model }X}{\text{\# of
    iterations in model }Y}  
\end{equation}
Like the simpler MCMC methods, the RJMCMC is \emph{guaranteed} to converge
on the correct value of $B_{XY}$ making it the `gold standard' of
Bayes Factor estimation.  And, like regular MCMCs, the convergence can be
very slow, so that in practice the Bayes Factor estimates can
be inaccurate. This is especially true when the trans-dimensional moves
involve many parameters, such as the 7 or 8 dimensional jumps that
are required to transition between models with differing numbers of
galactic binaries.

Figure 1 shows the output of a RJMCMC search of a simulated LISA
data stream containing the signal from a galactic binary with $q = {\dot f}T_{\rm obs}^2 =1$
and and observation time of $T_{\rm obs}=2$ years. The chain moved freely between the
7-dimensional model with no frequency evolution and the
8-dimensional model which included the frequency evolution.
\begin{figure}\label{RJ}
\includegraphics[angle=-90,width=0.5\textwidth]{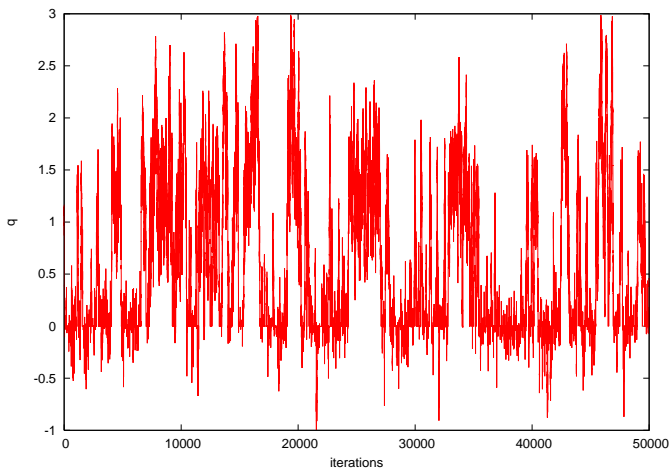}
\caption{50 000 iteration segment of an RJMCMC chain moving between
  models with and without frequency evolution.  This particular run was for a source
  with $q=1$ and SNR=10 and yielded $B_{XY}\sim 1$.}
\end{figure}

\subsection{Laplace Approximation}
A common approach to model selection is to approximate the model
evidence directly.  Working under the assumption that the PDF is
Gaussian, the integral in equation (\ref{evidence}) can be estimated by use of the
Laplace approximation.  This is accomplished by comparing the volume of
the models parameter space $V$ to that of the parameter uncertainty
ellipsoid $\Delta V$
\begin{equation}
p_X(s) \simeq p(\vec{\lambda}_{\rm MAP},X|s){\Big(}\frac{\Delta V_X}{V_X}{\Big)}.
\end{equation}
The uncertainty ellipsoid can be determined by calculating the
determinant of the Hessian ${\bf \mathcal{H}}$ of partial derivatives of the posterior with
respect to the model parameter evaluated at the MAP value for the parameters.
\begin{equation}
p_X(s) \simeq p(\vec{\lambda}_{\rm
  MAP},X|s)\frac{(2\pi)^{D/2}}{\sqrt{\det {\bf \mathcal{H}}}}
\end{equation}
The Fisher Information Matrix (FIM) ${\bf\Gamma}$ with
components 
\begin{equation}
\Gamma_{ij}\equiv(h,_i|h,_j)\ \  \text{where}\ \  h,_i\equiv
\frac{\partial h}{\partial \lambda^i}
\end{equation}
can be used as a quadratic  approximation to  ${\bf\mathcal{H}}$ yielding
\begin{equation}
p_X(s) \simeq p(\vec{\lambda}_{\rm
  MAP},X|s)\frac{(2\pi)^{D/2}}{\sqrt{\det {\bf \Gamma}}}
\end{equation}
We will refer to this estimate of the evidence as the Laplace-Fisher (LF) approximation.
The LF approximation breaks down if the priors have large gradients in the vicinity
of the MAP parameter estimates. The FIM estimates can also be poor if some of
the source parameters are highly correlated, or if the quadratic approximation fails.
In addition, the FIM approximation gets progressively worse as the SNR of the source decreases.

A more accurate (though more costly) method for estimating the evidence is the
Laplace-Metropolis (LM) approximation which employs the PDF as mapped out by the MCMC
exploration of the likelihood surface to estimate ${\bf \mathcal{H}}$~\cite{RJLMBIC}.  This
can be accomplished by fitting a minimum volume ellipsoid (MVE) to the $D$-dimensional
posterior distribution function.  The principle axes of the MVE lie in eigen-directions of the
distribution which generally do not lie along the parameter
directions.  Here we employ the {\tt MVE.jar} package which utilizes
a genetic algorithm to determine the MVE of the distribution and
returns the covariance matrix of the PDF~\cite{MVE}.  The determinant of the covariance
matrix can then be used to determine the evidence via
\begin{equation}
p_X(s) \simeq p(\vec{\lambda}_{\rm MAP},X|s)(2\pi)^{D/2}\sqrt{\det
  {\bf C}}.
\end{equation}
In the MCMC literature the LM approximation is generally considered to be second only to the
RJMCMC method for estimating Bayes Factors.
 
\subsection{Savage Dickie Density Ratio}
Both RJMCMC and LM require exploration of the posterior for each
model under consideration.  The Savage-Dickie (SD) approximation estimates the Bayes Factor
directly while only requiring exploration of the highest dimensional
space~\cite{bayes, SD}.  This approximation requires that two conditions are met: Model X must
be nested within Model Y (adding and subtracting parameters clearly
satisfies this condition) and the priors for any given model must be separable
(i.e. $p(\vec{\lambda})=p(\lambda^1) \times p(\lambda^2) \times \mathellipsis
\times p(\lambda^D)$) which is, to a good approximation, satisfied
in our example.  The Bayes Factor $B_{XY}$ is then calculated by comparing the weight of the
marginalized posterior to the weight of the prior distribution for the
`extra' parameter at the default, lower-dimensional, value for the
parameter in question.
\begin{equation}
B_{XY}(s) \simeq \frac{p(\lambda_0|s)}{p(\lambda_0)}
\end{equation}
It is interesting to note that if the above conditions
are precisely satisfied it can then be shown that this is an exact
calculation of $B_{XY}$ (assuming sufficient sampling of the PDF), as opposed to an approximation.

\subsection{Schwarz-Bayes Information Criterion}
All of the approximations discussed so far depend on the supplied priors
$p(\vec{\lambda})$. The Schwarz-Bayes Information Criterion (BIC) method is an approximation
to the model evidence which makes its own assumptions about the priors - namely that
they take the form of a multivariate Gaussian with covariance matrix derived from the
Hessian ${\bf \mathcal{H}}$~\cite{RJLMBIC, BIC}. The BIC estimate for the
evidence is then
\begin{equation}
\ln p_X(s) \simeq \ln p(\vec{\lambda}_{\rm MAP},X|s)-\frac{D_X}{2}\ln
N_{\rm eff}
\end{equation}
where $D_X$ is the dimension of model $X$ and $N_{\rm eff}$ is the
\emph{effective} number of samples in the data. For our tests we defined
$N_{\rm eff}$ to be the number of data points required to return a (power)
signal-to-noise ratio of ${\rm SNR}^2 - D$, where SNR is the signal-to-noise
one gets by summing over the entire LISA band. This choice was motivated by
the fact that the variance in SNR$^2$ is equal to $D^2$, so $N_{\rm eff}$
accounts for the data points that carry significant weight in the model
fit. The BIC estimate has the advantage of being very easy to calculate,
but is generally considered less reliable than the other techniques
we are using.

\section{Case Study}
To compare the various approximations to the Bayes Factor we simulated a
`typical' galactic binary.  The injected parameters for our test source can be found in table
\ref{tab2}.  Since ${\dot f} \sim f^{11/3}$, higher frequency sources are more
likely to have a measurable $\dot{f}$. On the other hand, the number of
binaries per frequency bin falls of as $\sim f^{-11/3}$, so high frequency
systems are fairly rare. As a compromise, we selected a system with a GW
frequency of 5 mHz. To describe the frequency evolution we introduced the
dimensionless parameter
\begin{equation}
q\equiv \dot{f} T_{\rm obs}^2,
\end{equation}
which measures the change in the Barycenter GW frequency in units of
$1/T_{\rm obs}$ frequency bins. For $q \gg 1$ it is reasonable to believe that
a search algorithm will have no difficulty detecting the frequency shift. 
Likewise, for $q \ll 1$ it is unlikely that the frequency evolution can be
detected (at least for sources with modest SNR). Therefore we have selected
$q \sim 1$ to test the model selection techniques. Achieving $q=1$
for typical galactic binaries at 5 mHz requires observation times of
approximately two years. A range of SNRs were explored by varying the
distance to the source.

\begin{table}[h]
\caption{Source parameters}
\begin{tabular}{|c|c|c|c|c|c|c|c|}
\hline
$f$ (mHz) & $\cos{\theta}$ & $\phi$ (deg) & $\psi$ (deg) & $\cos{\iota}$ &
$\varphi_0$ (deg) & q & $T_{\rm obs}$ (yr)\\
\hline 
5.0 & 1.0 & 266.0 & 51.25 & 0.17 & 204.94 &
1 & 2 \\
\hline  
\end{tabular}
\label{tab2}
\end{table}

We can rapidly calculate accurate waveform templates using the fast-slow decomposition
described in the Appendix. Our waveform algorithm has been used in the second round of
Mock LISA Data Challenges~\cite{MLDC2} to simulate the response to a galaxy containing
some 26 million sources. The simulation takes just a few hours to run on a single 2 GHz
processor.

We simulated a 1024 frequency bin snippet of LISA data around 5 mHz that included the injected
signal and stationary Gaussian instrument noise. The Markov chains were initialized at
the injected source parameters as the focus of this study is the statistical character
of the detection, and not the initial detection (a highly efficient solution to the
detection problem is described in Ref.~\cite{BAM}). We used uniform priors for all of the
parameters, with the angular parameters taking their standard ranges. We took the frequency
to lie somewhere within the frequency snippet, and $\ln A$ to be uniform across
the range $\frac{1}{2}\ln(S_n/(2T))$ and $\frac{1}{2}\ln(1000 S_n/(2T)$, which roughly
corresponds to signal SNRs in the range 1 to 1000. We took the frequency evolution parameter
$q$ to be uniformly distributed between -3 and 3 and adopted $q=0$ as
the default value when operating under the 7-dimensional model. In reality, astrophysical considerations
yield a very strong prior for $q$ (see Section \ref{astro_p}) that will significantly impact
model selection. We decided to use a simple uniform prior to compare the various approximations
to the Bayes Factor, before moving on to consider the effects of the astrophysical prior in
Section \ref{astro_p}.

The choice of proposal distribution $q(\vec{x}|\vec{y})$ from which to draw new parameter
values has no effect on the asymptotic form of the recovered PDFs, but the choice is
crucially important in determining the rate of convergence to the stationary distribution.
We took $q(\vec{x}|\vec{y})$ to be a multivariate Gaussian with covariance matrix given
by the inverse of the FIM. In addition to the source parameters we included two additional
parameters, $k_A$ and $k_E$, that describe the noise levels in the A and E data channels:
\begin{eqnarray}
S_n^A(f) &=& k_A S_n(f) \nonumber\\
S_n^E(f) &=& k_E S_n(f).
\end{eqnarray}
In a given realization of the instrument noise $k_A$ and $k_E$ will differ from
unity by some random amount $\delta$. The quantity $\delta$ will have a Gaussian
distribution with variance $\sigma^2 = 1/N$, where $N$ is the number of frequency
bins being analyzed. The likelihood $p(s|\vec{\lambda})$ can then be written as
\begin{equation}
p(s|\vec{\lambda})=C' \exp\Big[-\frac{1}{2}\Big(s-h\Big |
  s-h\Big) -  N\ln{(k_A k_E)}\Big]
\end{equation}
where the constant $C'$ is independent of the signal parameters $\vec{\lambda}$
{\em and} the noise parameters $k_A$ and $k_E$. We explored the
noise level estimation capabilities of our MCMC algorithm by
starting $k_A$ and $k_E$ far from unity. As can be seen in Figure 2
the chain quickly identified the correct noise level.

\begin{figure}[h]\label{nl}
\includegraphics[angle=-90,width=0.5\textwidth]{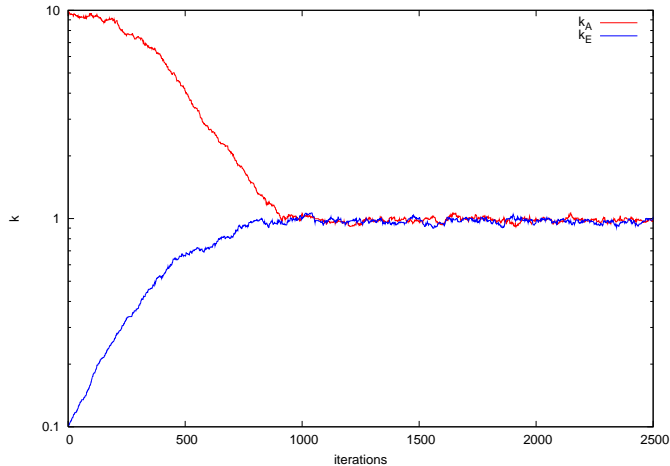}
\caption{Demonstration of the MCMC algorithm's rapid determination of the
injected noise level. The parameters $k_A$ and $k_E$ were initialized at 10 and
0.1 respectively.}
\end{figure}

\section{Comparison of Techniques}\label{tech}

We compared the Bayes Factor estimates obtained using the various methods in two
ways. First, we fixed the frequency derivative of the source at $q=1$ and varied
the SNR between 5 and 20 in unit increments. Second, we fixed the signal to noise
ratio at ${\rm SNR} =12$ and varied the frequency derivative of the source.

\begin{figure}[h]\label{b_snr}
\includegraphics[width=0.5\textwidth]{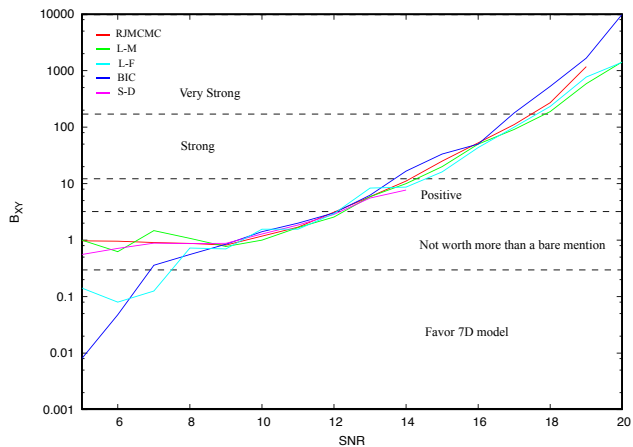}
\caption{Plot of the Bayes Factor estimates as a function of SNR for each of the
approximation schemes described in the text.}
\end{figure}

The results of the first test are shown in Figure 3
We see that all five methods agree very well for ${\rm SNR} > 7$. As expected,
the Laplace-Metropolis and Savage-Dicke methods provide the best approximation to
the ``Gold Standard'' RJMCMC estimate, showing good agreement all the way down to
${\rm SNR} = 5$. Most importantly, all five methods agree on when the 8-dimensional
source model is favored over the 7-dimensional model, placing the transition point
at ${\rm SNR} \simeq 12.2$. To get a rough estimate for the numerical
error in the various Bayes Factor estimates we repeated the SNR$=15$ case 10 times
using different random number seeds. We found that
the numerical error was enough to account for any quantitative differences between
the estimates returned by the various approaches. 

It is interesting to compare the Bayesian model selection results to the
frequentist ``3-$\sigma$'' rule for positive detection:
\begin{equation}
\vert {\bar q} \vert > 3 \sigma_q ,
\end{equation}
where ${\bar q}$ is the MAP estimate for the frequency change and $\sigma_q$ is
the standard deviation in $q$ as determined by the FIM. For the source under
consideration we found the ``3-$\sigma$'' rule to require ${\rm SNR} \simeq 13$ for
a detection, in good agreement with the Bayesian analysis. This lends support to
Seto's~\cite{seto} earlier FIM based study of the detectability of the frequency
evolution of galactic binaries, but we should caution that the literature is
replete with examples where the ``3-$\sigma$'' rule yields results in disagreement
with Bayesian model selection and common sense~\cite{lindley}.

\begin{figure}[h]\label{b_q}
\includegraphics[width=0.5\textwidth]{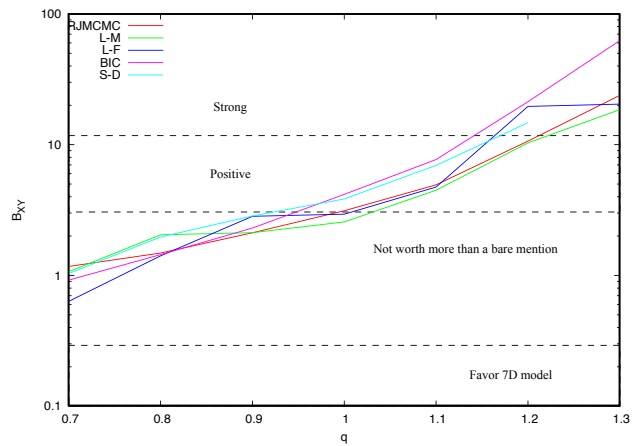}
\caption{Plot of the Bayes Factor estimates as a function of $q$ for each of the
approximation schemes described in the text. The signal to noise ratio was held
fixed at ${\rm SNR} = 12$.}
\end{figure}

The results of the second test are displayed in Figure 4
In this case
all five methods produced results that agree to within numerical error.

While the results shown here are for a particular choice of source parameters,
we found similar results for other sets of source parameters. In general
all five methods for estimating the Bayes Factor gave consistent results for
signals with ${\rm SNR} > 7$. One exception to this general trend were sources
with inclinations close to zero, as then the PDFs tend to be highly non-gaussian.
The Laplace-Metropolis and Laplace-Fisher approximations suffered the most
in those cases.

\section{Astrophysical Priors}\label{astro_p}

Astrophysical considerations lead to very strong priors for the frequency evolution
of galactic binaries. The detached systems, which are expected to account for the
majority of LISA sources, will evolve under gravitational radiation reaction in
accord with the leading order quadrapole formula:
\begin{equation}\label{rr}
{\dot f} = \frac{3 (8\pi)^{8/3}}{40} f^{11/3} {\cal M}^{5/3} \, ,
\end{equation}
where ${\cal M}$ is the chirp mass. Contact binaries undergoing stable mass transfer
from the lighter to the heavier component are driven to longer orbital periods by
angular momentum conservation. The competition between the effects of mass transfer
and gravitational wave emission lead to a formula for ${\dot f}$ with the same
frequency and mass scaling as (\ref{rr}), but with the opposite sign and a slightly
lower magnitude~\cite{nyz04}.

Population synthesis models, calibrated against observational data, yield predictions
for the distribution of chirp masses ${\cal M}$ as a function of orbital frequency.
These distributions can be converted into priors on $q$. 
In constructing such priors one should also fold in observational selection
effects, which will favor systems with larger chirp mass (the GW amplitude scales as
${\cal M}^{5/6}$). To get some sense of how such priors will affect the model
selection we took the chirp mass distribution for detached systems at $f\sim 5$ mHz
from the population synthesis model described in Ref.~\cite{nyz01}, (kindly provided
to us by Gijs Nelemans), and used (\ref{rr}) to construct the prior on $q$ shown
in Figures 5 and 6 (observation selection effects were ignored). The prior has
been modified slightly to give a small but no-vanishing weight to sources with $q=0$.
The astrophysically motivated prior has a very sharp peak at $q=0.64$, and we use
this value when fixing the frequency derivative for the 7-dimensional model. 

\begin{figure}[h]\label{astroSNR}
\includegraphics[angle=0,width=0.5\textwidth]{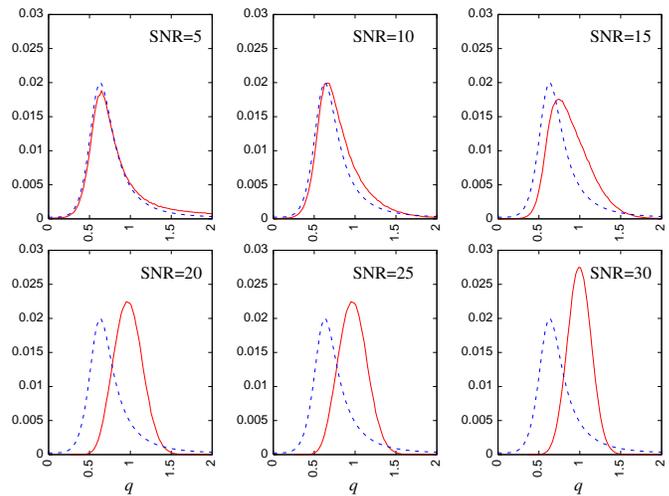}
\caption{Comparison between astrophysically motivated prior distribution of $q$ for $f=5$ mHz
and $T_{\rm obs}=2$ years (dashed, blue) to marginalized PDF (solid, red) for sources injected
with $q=1$ and SNRs varying from $5$ to $30$.}
\end{figure}

To explore the impact on model selection when such a strong prior has
been adopted we simulated a source with $q=1$ and varied the SNR.  The RJMCMC
algorithm was applied using chains of
length $10^7$ in conjunction with a fixed 8-dimensional MCMC (also
allowed to run for $10^7$ iterations) in order to compare the RJMCMC
results with the Savage-Dickie density ratio.

\begin{table}[h]
\begin{tabular}{|c|c|c|}
\hline
SNR & $B_{XY}$ (SD) & $B_{XY}$ (RJMCMC)\\
\hline 
5 & 0.926 & 1.015\\
10 & 0.977 & 0.996\\
15 & 0.749 & 0.742\\
20 & 0.427 & 0.427\\
25 & 0.176 & 0.177\\
30 & 0.060 & 0.056\\
\hline  
\end{tabular}
\label{tab3}
\caption{Savage-Dickie density ratio estimates of $B_{XY}$ for sources
with $q=1$ and SNRs varying from 5 to 30.  Comparisons with RJMCMC
explorations of the same data set show excellent agreement between the
two methods.}
\end{table}

The results of this first exploration are shown in Figure 5.  We found that for
${\rm SNR} < 15$ the marginalized PDF very closely
resembled the prior distribution.  This demonstrates that the information content
of the data is insufficient to change our prior belief about the value
of the frequency derivative.  As the SNR increased, however, the
PDF began to move away from the prior until we reached SNR=30 when
the astrophysical prior had a negligible effect on the shape of
the posterior, signaling confidence in the quoted measurement of
$q$.    This qualitative assessment of model preference is strongly
supported by the Bayes factor estimation made by the RJMCMC
algorithm as can be seen in Table III.  It should also be noted that
the excellent agreement between the RJMCMC and S-D estimates for
Bayes factor $B_{XY}$. Both methods indicate that for the chosen value
of $q=1$, the signal-to-noise needs to exceed ${\rm SNR} \sim 25$ for the 8-dimensional model
to be favored. This is in contrast to the case discussed earlier where a
uniform prior was adopted for the frequency derivative, and the model
selection methods began showing a preference for the 8-dimensional
model around SNR=12.

\begin{figure}[h]\label{astroq}
\includegraphics[angle=0,width=0.5\textwidth]{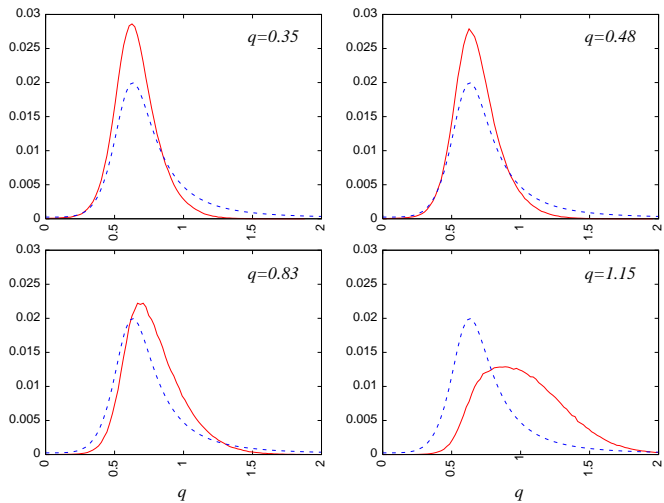}
\caption{Marginalized PDF (solid, red) for fixed SNR=15 injected sources
  with $q$ corresponding to FWHM and FWQM of the astrophysical prior (dashed, blue)}
\end{figure}

\begin{table}[h]
\begin{tabular}{|c|c|c|}
\hline
$q$ & $B_{XY}$ (SD) & $B_{XY}$ (RJMCMC)\\
\hline 
0.35 & 1.412 & 1.414\\
0.48 & 1.381 & 1.388\\
0.83 & 1.059 & 1.052\\
1.15 & 0.432 & 0.428\\
\hline  
\end{tabular}
\label{tab4}
\caption{Savage-Dickie and RJMCMC density ratio estimates of $B_{XY}$ for sources
with SNR=15 and $q$ at FWHM and FWQM of astrophysical prior}
\end{table}

Figure 6 shows the impact of the astrophysically motivated prior when the SNR was held at 15 and
four different injected values for $q$ were adopted, corresponding to
the full width at half maximum (FWHM)
and full width at quarter maximum (FWQM) of the prior distribution.
The Bayes factors listed in Table IV indicate
that for modestly loud sources with
SNR=15 the model selection techniques do not favor updating our
estimate of the frequency derivative until the frequency derivative exceeds $q = 1.2$.

\section{Discussion}

We have found that the several common methods for estimating Bayes Factors
give good agreement when applied to the the model selection problem of
deciding when the data from the LISA observatory can be used to detect
the orbital evolution of a galactic binary. The methods studied require
varying degrees of effort to implement and calculate, and although
found to be accurate in this test case, it is clear that some of these
methods would be inappropriate
approximations for more physically relevant examples.

If a RJMCMC algorithm is used as the sole model selection technique,
the resistance of the algorithm to change dimension, especially when
making multi-dimensional jumps, can result in invalid model selection
unless the chains are run for a very large numbers of steps.
In the examples we studied the transdimensional jumps only had to span
one dimension, and our basic RJMCMC algorithm performed well. However,
a more sophisticated implementation, using {\it e.g.} rejection sampling
or coupled chains, will be required to select the number of sources,
as this requires jumps that span seven or more dimensions. 

The Laplace-Metropolis method for approximating the model evidence is
more robust than the commonly used Fisher Information Matrix
approximation of the Hessian of the PDF.  Implementing an LM evidence
estimation is a somewhat costly because of the need to fit the posterior
to a minimum volume ellipsoid.

The Savage-Dickie approximation is more economical than the RJMCMC or LM
methods, but is limited by the requirement that the
competing models \emph{must} be nested.

The Bayes Information Criterion approximation to the evidence is by far the
cheapest to implement, and is able to produce reliable results when the
SNR is high. It has therefore shown the most promise as an `on the fly' model
determination scheme.  More thorough (and therefore more costly)
methods such as RJMCMC and LM could then be used to refine the
conclusions initially made by the BIC.

Our investigation using a strong astrophysical prior indicated that
the gravitational wave signals will need to have high signal-to-noise (${\rm SNR} > 25$),
or moderate signal-to-noise (${\rm SNR} > 15$) and frequency derivatives far from the
peak of the astrophysical distribution,
in order to update our prior belief in the value of the
frequency derivative. In other words, the frequency derivative will only
been needed as a search parameter for a small number of bright high frequency sources.

\section*{Acknowledgments}
This work was supported by NASA Grant NNG05GI69G. We are most grateful to Gijs Nelemans for
providing us with data from his population synthesis studies.

\section*{Appendix A}\label{ApA}
To leading order in the eccentricity, $e$, the Cartesian coordinates of
the $i^{\rm th}$ LISA spacecraft are given by~\cite{CR03}
\begin{eqnarray} \label{keporb}
x_i(t) &=& R \cos(\alpha) + \frac{1}{2} e R  \Big( \cos(2\alpha-\beta_i) -
3\cos(\beta) \Big) 
\nonumber\\
y_i(t) &=& R \sin(\alpha) + \frac{1}{2} e R  \Big( \sin(2\alpha-\beta_i) -
3\sin(\beta) \Big) \nonumber\\
z_i(t) &=& -\sqrt{3} e R  \cos(\alpha-\beta_i)  \; .
\end{eqnarray}
In the above $R = 1$ AU, is the radial distance of the guiding center,
$\alpha = 2\pi f_m t + \kappa$ is the orbital
phase of the guiding center, and $\beta_i = 2\pi (i-1)/3 + \lambda$
($i=1,2,3$) is the relative phase of the spacecraft within the
constellation.  The parameters $\kappa$ and $\lambda$ give the initial
ecliptic longitude and orientation of the constellation. The distance
between the spacecraft is $L = 2 \sqrt{3} e R$. Setting $e=0.00985$
yields $L = 5 \times 10^9$ m.

An arbitrary gravitational wave traveling in the $\hat{k}$ direction can
be written as the linear sum of two independent polarization states
\begin{equation}
{\bf h}(\xi) = h_+(\xi) {\bf \varepsilon}^+ + h_\times(\xi)
{\bf \varepsilon}^\times
\end{equation}
where the wave variable $\xi = t - \hat{k} \cdot {\bf x}$ gives the
surfaces of constant phase. The polarization tensors can be expanded in
terms of the basis tensors ${\bf e}^+$ and
${\bf e}^\times$ as
\begin{eqnarray}
{\bf \varepsilon}^+ &=& \cos(2\psi) {\bf e}^+ -\sin(2\psi) {\bf e}^\times
\nonumber \\
{\bf \varepsilon}^\times &=& \sin(2\psi) {\bf e}^+ + \cos(2\psi) 
{\bf e}^\times \; ,
\end{eqnarray}
where $\psi$ is the polarization angle and 
\begin{eqnarray}
{\bf e}^+ &=& \hat{u} \otimes \hat{u} - \hat{v} \otimes \hat{v}
\nonumber\\
{\bf e}^\times &=& \hat{u} \otimes \hat{v} + \hat{v} \otimes \hat{u} \,.
\end{eqnarray}
The vectors $(\hat{u}, \hat{v}, \hat{k})$ form an orthonormal triad
which may be expressed as a function of
the source location on the celestial sphere
\begin{eqnarray}\label{basis}
\hat{u} &=& \cos\theta\cos\phi\,\hat{x} + \cos\theta\sin\phi\,\hat{y} -
\sin\theta\,\hat{z} \nonumber\\
\hat{v} &=& \sin\phi\,\hat{x} - \cos\phi\,\hat{y} \nonumber\\
\hat{k} &=& -\sin\theta\cos\phi\,\hat{x} - \sin\theta\sin\phi\,\hat{y} -
\cos\theta\,\hat{z} \,.
\end{eqnarray}
For mildly chirping binary sources we have
\begin{equation}
{\bf h}(\xi) = \Re \Big[ 
\left(A_+ {\bf \varepsilon}^+ 
+ e^{i 3\pi/2}A_\times {\bf \varepsilon}^\times\right) e^{i\Psi(\xi)}\Big]
\end{equation}
where
\begin{eqnarray}
&&A_+ = \frac{2 {\cal M}(\pi f)^{2/3}}
{D_L}\left(1 + \cos^2 \iota \right) \nonumber \\
&&A_\times = -\frac{4{\cal M}(\pi f)^{2/3}} {D_L}\cos \iota \,.
\end{eqnarray}
Here ${\cal M}$ is the chirp mass, $D_L$ is the luminosity distance
and $\iota$ is the inclination of the binary to the line of sight.
Higher post-Newtonian corrections, eccentricity of the orbit, and spin
effects will introduce additional harmonics. For chirping sources the
adiabatic approximation requires that the
frequency evolution $\dot f$ occurs on a timescale long compared to
the light travel time in the interferometer: $f / \dot f \ll L$.
The gravitational wave phase can be approximated as
\begin{equation}
\Psi(\xi) = 2\pi f_0 \xi + \pi \dot f_0 \xi^2 + \varphi_0 \, ,
\end{equation}
where $\varphi_0$ is the initial phase. The instantaneous
frequency is given by $2 \pi f = \partial_t \Psi$:
\begin{equation}
f = (f_0 + \dot f_0 \xi)(1-\hat{k}\cdot{\bf v}) \, .
\end{equation}
The general expression for the path length variation caused by a
gravitational wave involves an integral in $\xi$ from $\xi_i$ to
$\xi_f$. Writing $\xi = \xi_i + \delta \xi$ we have
\begin{equation}
\Psi(\xi) \simeq 2\pi (f_0 + \dot f_0 \xi_i) \delta \xi + {\rm const}\, .
\end{equation}
Thus, we can treat the wave as having fixed frequency
$f_0 + \dot f_0 \xi_i$ for the purposes of the integration, and
then increment the frequency forward in time in the final expression~\cite{RAA}.
The path length variation is then given by~\cite{CR03,RAA}
\begin{equation}\label{delRA}
\delta\ell_{ij}(\xi) = L  \Re \Big[ {\bf d}(f,t,\hat{k}):{\bf h}(\xi)\Big]\, ,
\end{equation}
where ${\bf a}:{\bf b} = a^{ij}b_{ij}$.
The one-arm detector tensor is given by
\begin{equation}
{\bf d}(f,t,\hat{k}) = \frac{1}{2}\Big(\hat{r}_{ij}(t) \otimes
  \hat{r}_{ij}(t)\Big) {\mathcal T}(f,t,\hat{k}) \, ,
\end{equation}
and the transfer function is
\begin{eqnarray}
{\mathcal T}(f,t,\hat{k}) &=& \textrm{sinc} \left( \frac{f}{2
f_*}\Big(1 - \hat{k} \cdot \hat{r}_{ij}(t)\Big) \right) \nonumber\\
&& \times \exp \left( i \frac{f}{2 f_*}\Big(1 - \hat{k} \cdot
    \hat{r}_{ij}(t) \Big) \right) \, ,
\end{eqnarray}
where $f_* = 1/(2\pi L)$ is the transfer frequency and
$f = f_0 + \dot f_0 \xi$. The expression
can be attacked in pieces. It is useful to define the quantities
\begin{eqnarray}
d^{+}_{ij}(t) &\equiv& \left( \hat{r}_{ij}(t) \otimes \hat{r}_{ij}(t)
\right) : {\bf e}^{+} \\
d^{\times}_{ij}(t) &\equiv& \left( \hat{r}_{ij}(t) \otimes
\hat{r}_{ij}(t) \right)  :{\bf e}^{\times} \,.
\end{eqnarray}
and  $y_{ij}(t) = \delta\ell_{ij}(t)/(2L)$. Then
\begin{equation}\label{main}
y_{ij}(t) = \Re \left[ y^{\rm slow}_{ij}(t)  e^{2 \pi i f_0 t} \right],
\end{equation}
where
\begin{eqnarray}
&& y^{\rm slow}_{ij}(t) = \frac{{\mathcal T}(f,t,\hat{k})}{4}\left(\left(d^+_{ij}(t)(A_+(t)\cos(2\psi)
 \right. \right. \nonumber \\
&& \quad +e^{3\pi i/2}A_\times(t)\sin(2\psi)) \nonumber \\
&&  \quad + d^\times_{ij}(t)(e^{3\pi i/2}A_\times(t)\cos(2\psi) \nonumber \\
&& \quad  \left. \left. -A_+(t)\sin(2\psi))\right)
e^{(\pi i \dot f_0 \xi^2 + i\varphi_0 - 2 \pi i f_0 \hat{k}\cdot {\bf x} )}\right)
\end{eqnarray}
It is a simple exercise to derive explicit expressions for the
antenna functions and the transfer function appearing in $y^{\rm slow}_{ij}(t)$
using (\ref{keporb}) and (\ref{basis}).

In the Fourier domain the response can be written as
\begin{equation}\label{fast_slow}
y_{ij}(t) = \Re \Big[ \left(\sum_n a_n e^{2\pi i n
t/T_{\rm obs}}\right)
e^{2 \pi i f_0 t} \Big]\, ,
\end{equation}
where the coefficients $a_n$ can be found by a numerical FFT of the slow terms
$y^{\rm slow}_{ij}(t)$. Note that the sum over $n$ should extend over both negative and
positive values. The number of time samples needed in the FFT will depend on $f_0$
and $\dot f_0$ and $T_{\rm obs}$, but is less than $2^9=512$ for any galactic sources
we are likely to encounter when $T_{\rm obs} \leq {\rm 2 yr}$. The
bandwidth of a source can be estimated as
\begin{equation}
B = 2 \left( 4 + 2\pi f_0 R \sin (\theta) \right) f_m 
+ \dot f_0 T_{\rm obs} \, .
\end{equation}
The number of samples should exceed $2B T_{\rm obs}$. The Fourier transform of the fast term can be
done analytically:
\begin{equation}\label{fast_ft}
e^{2 \pi i f_0 t} = \sum_m b_m e^{2\pi i m t/T_{\rm obs}}
\end{equation}
where
\begin{equation}
b_m = T_{\rm obs} \, {\rm sinc} (x_{m}) e^{i x_{m}}
\end{equation}
and
\begin{equation}
x_{m} = f_0 T_{\rm obs} - m \, .
\end{equation}
The cardinal sine function in (\ref{fast_ft}) ensures that
the Fourier components $b_m$ away from resonance, $x_{m} \approx 0$,
are quite small. It is only necessary to keep $\sim 100 \rightarrow 1000$ terms either
side of $p = [f_0 T_{\rm obs}]$, depending on how bright the source is, and how far
$f_0 T_{\rm obs}$ is from an integer. We now have
\begin{equation}\label{combined}
y_{ij}(t) = \Re \Big[ \left(\sum_j c_j e^{2\pi i j
t/T_{\rm obs}}\right) \Big]\, ,
\end{equation}
where
\begin{equation}
c_j = \sum_n a_n b_{j-n} \, .
\end{equation}
The final step is to ensure that our Fourier transform yields a real
$y_{ij}(t)$. This is done by setting the final answer for
the Fourier coefficients equal to $d_j = (c_j + c^*_{-j})/2$. 
But since $x_{m}$ never hits resonance for positive
$j$ (we are not interested in the negative frequency components $j<0$),
we can neglect the second term and simply write $d_j = c_j /2$.

Basically what we are doing is hetrodyning the signal to the base
frequency $f_0$, then Fourier transforming the slowly evolving
hetrodyned signal numerically. We then convolve these Fourier
coefficients with the analytically derived Fourier coefficients
of the carrier wave.

The Michelson type TDI variables are given by
\begin{eqnarray}\label{alpha}
X(t)&=&y_{12}(t-3L)-y_{13}(t-3L) +y_{21}(t-2L) \nonumber \\
&& - y_{31}(t-2L) +y_{13}(t-L)-y_{12}(t-L) \nonumber \\
&& +y_{31}(t) - y_{21}(t), \\
Y(t)&=&y_{23}(t-3L)-y_{21}(t-3L) +y_{32}(t-2L) \nonumber \\
&& - y_{12}(t-2L) +y_{21}(t-L)-y_{23}(t-L)\nonumber \\
&& +y_{12}(t) - y_{32}(t), \\
Z(t)&=&y_{31}(t-3L)-y_{32}(t-3L) +y_{13}(t-2L) \nonumber \\
&& - y_{23}(t-2L) +y_{32}(t-L)-y_{31}(t-L) \nonumber \\
&& +y_{23}(t) - y_{13}(t).
\end{eqnarray}
Note that in the Fourier domain
\begin{eqnarray}
X(f) &=& \tilde y_{12}(f) e^{-3 i f/f*} - \tilde y_{13}(f) e^{- 3 i f/f*}
+ \tilde y_{21}(f)e^{-2 i f/f*} \nonumber \\
&& - \tilde y_{31}(f)e^{-2 i f/f*} +  \tilde y_{13}(f) e^{-i f/f*}
- \tilde y_{12}(f) e^{- i f/f*} \nonumber \\
&& + \tilde y_{31}(f) - \tilde y_{21}(f) \, .
\end{eqnarray}
This saves us from having to interpolate in the time domain. We just
combine phase shifted versions of our original Fourier transforms.

\end{document}